# 3D track reconstruction of low-energy electrons in the MIGDAL low pressure optical time projection chamber

**E. Tilly,**[a] **M. Handley**[b,c] **on behalf of the MIGDAL collaboration**[c]

[a] *Department of Physics and Astronomy, University of New Mexico,*
 *210 Yale Blvd NE Albuquerque, NM 87106, United States*
 E-mail: tillyeg01@unm.edu

[b] *Cavendish Laboratory, University of Cambridge,*
 *J J Thomson Avenue Cambridge CB3 0HE, United Kingdom*

[c] *Particle Physics Department, STFC Rutherford Appleton Laboratory*
 *Didcot OX11 0QX, United Kingdom*

ABSTRACT: We demonstrate three-dimensional track reconstruction of electrons in a low pressure (50 Torr) optical TPC consisting of two glass GEMs with an ITO strip readout in $CF_4$ and $CF_4/Ar$ mixtures. The reconstructed tracks show a variety of event topologies, including short tracks from photoelectrons induced by $^{55}Fe$ 5.9 keV X-rays and long tracks from gamma ray interactions and beta decays. Algorithms for event identification and track ridge detection are discussed as well as multiple methods for integrating information from the camera image and ITO waveforms with the goal of full 3D reconstruction of the track.

KEYWORDS: Data processing methods; Gaseous imaging and tracking detectors; Optical detector readout concepts; Time projection chambers.

**Contents**

## 1. Introduction

The Migdal effect is defined as the ionization of an atom from a collision between its atomic nucleus and a neutral particle [1-3]. This effect has been invoked by a number of dark matter experiments to extend their low mass reach up to two orders of magnitude [4-11]. But, because this effect has never been experimentally observed, the MIGDAL collaboration has begun an effort to directly detect and characterize it using a low-pressure gas optical time projection chamber (TPC) described in [1]. Specifically, the experiment is designed to make a direct observation of the characteristic Migdal topology, namely a nuclear recoil sharing an interaction vertex with an electron. Because dark matter experiments use this effect in a very low energy regime, it is imperative that this experiment can resolve and reconstruct low energy tracks[1].

    The MIGDAL detector is a low pressure Optical Time Projection Chamber (OTPC), filled to 50 Torr with $CF_4$ gas. Track ionisation within the 3cm drift region is drifted to a double glass-GEM where it is amplified and collected by an indium tin oxide (ITO) strip readout plane. The scintillation light produced during amplification is read out both by a scientific Hamamatsu ORCA-Fusion, C14440 CMOS camera and a VUV-sensitive, 3-inch Hamamatsu R11410 photomultiplier tube (PMT), which is also sensitive to the primary scintillation of the track. The camera images a 2D projection of the particle tracks in the x-y plane, and the information is combined with the 2D x-z projection measured by the ITO strips to reconstruct tracks in 3D. Measurements of the primary and secondary scintillation from the PMT are used to reconstruct the absolute depth of the track, but its inclusion in 3D reconstruction is not discussed here. The primary goal of this paper is to describe how information from the ITO strip readout can be combined with camera image data to get accurate 3D reconstructions. This process is done using two methods, described in the following section.

---

[1] Though the proposed observations will still be higher energy than those invoked by dark matter experiments [3].

## 2. 3D Reconstruction

### 2.1 Voxelisation

The first method reconstructs the full 3D energy deposition of a given event inside the detector. Because the camera image and the ITO readouts are projections of the event in two perpendicular planes, all that is required is to combine these to reconstruct the 3D track. This is done by performing the following operation:

$$R_{i,j,k} = ITO_{i,j} \times Im_{i,k}$$

where $R_{i,j,k}$ is the (i,j,k) voxel of the reconstruction, $ITO_{i,j}$ is the (i,j) sample of the ITO strip readout and $Im_{i,k}$ is the (i,k) pixel of the image. Before this operation can be done, some preprocessing on the images and strip readouts is needed.

The camera images, in their raw form, are quite noisy and, oftentimes, the detector's GEM holes are visible in the images. Thus, it is important to remove these artifacts and isolate the track as much as possible to get a good 3D reconstruction. The first step, much like in astronomy, is to correct for overall offsets (pedestals) in the camera pixels by performing a dark subtraction of the raw image. Once this has been done, the signal to noise ratio (SNR) is enhanced by multiplying a Gaussian kernel to the Fourier transform of each image to emphasize lower frequencies (containing the track) and suppress higher frequencies (i.e. the noise and GEM holes) within the image. Finally, to isolate the tracks within each image, a mask is applied which keeps all pixels with intensities above a set threshold, setting the rest equal to zero. The strip readout, which intrinsically has a relatively high SNR, is similarly masked such that the tracks are kept and everything else is rejected.

Because the two data types come with very different granularities (camera image is ~70 microns/pixel and strip readout has a pitch of 833 microns) the camera image must either be binned to match the low granularity of the strip readout or the strip readout must be artificially upscaled to match the fine granularity of the camera image.

Once the preprocessing is complete, the reconstruction algorithm as described by the equation above is applied, resulting in a voxelized representation of the tracks; an example is shown in Figure 1. This method is particularly handy in describing the full spatial distribution of energy deposited along a given track.

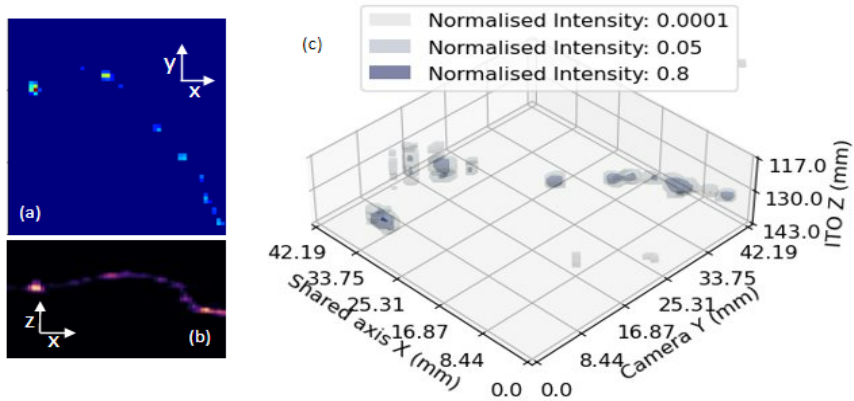

**Figure 1.** Example of processed and down scaled camera image (a), processed electronic readout (b) and voxelised reconstruction (c). Each of the blue bubbles in the voxelised reconstruction represents a 3D contour of the normalized intensity of the voxelization of the track.

**2.2 RidgeFinder**

In order to fully characterize the tracks in the MIGDAL experiment, the reconstruction must also be able to extract more detailed features of the tracks such as particle identification (PID), length, dE/dx, initial direction, etc. Thus, a second algorithm has been developed to obtain this information by reconstructing the best-guess particle trajectory.

As in the voxelised approach, the raw image is preprocessed by subtracting a mean dark image and removing noise and GEM holes by applying a Gaussian kernel in Fourier space. The resulting image undergoes further processing, not used in the voxelised method, to correct for electron diffusion that occurs during the drift and amplification stages. This is done using the iterative Lucy-Richardson deconvolution algorithm, which uses a Gaussian point spread function with a σ that matches the average diffusion in the TPC.

With preprocessing completed, the track's 2D ridgeline is extracted from the image using an algorithm developed by Carsten Steger [12]. This was originally developed for identifying ridge structures in biological imagery but we have found it also works well for reconstructing particle tracks. The algorithm calculates the discrete Hessian matrix for each pixel in the image then uses the highest eigenvalues to identify points of maximal curvature. These are linked using the eigenvectors to form a 2D ridgeline, which is then smoothed with a spline.

The time of the peak amplitude of the deconvolved ITO signal in a given strip is converted to a relative z position using the drift velocity. In the absence of topological ambiguity, this gives an approximation for the center of the charge distribution in z along a given strip (x). These points are then correlated with the ridgeline found from the camera image using the common coordinate, x, to form the 3D reconstruction. An example of this is shown in Figure 2.

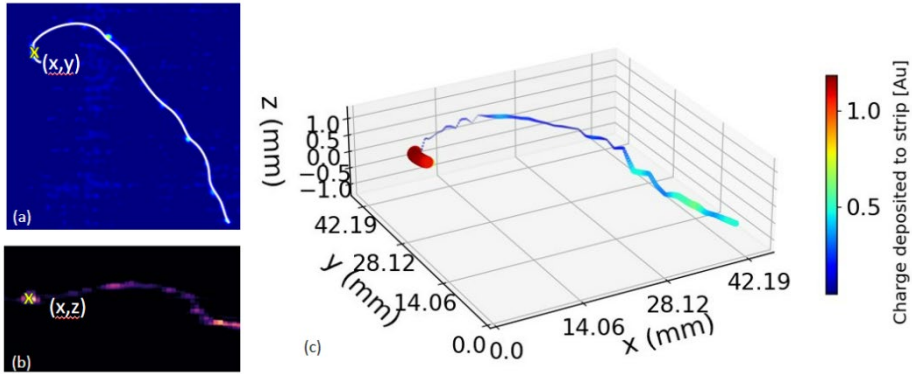

**Figure 2.** Example of camera image with a 2D ridgeline fitted to it (a), corresponding strip readout point (b) and 3D ridgeline reconstruction (c). For each point of the 3D ridge, the associated intensity, as found from the ITO, has been included and is represented by color.

**2.3 Extent of Reconstruction**

It is vital that the MIGDAL experiment can successfully reconstruct low energy electron tracks down to 5 keV, a conservative threshold energy as described in [1]. Using the methods described here, we have demonstrated this[2] using X-rays from $^{55}$Fe to produce both 5.9 keV and 2.9 keV electron tracks, the latter from the Ar escape peak, as shown in Figure 3. In addition, our method has also been successfully applied to faint, low dE/dx betas (see Figures 1 and 2). The results show that our techniques are capable of reconstructing events as expected in the MIGDAL detector.

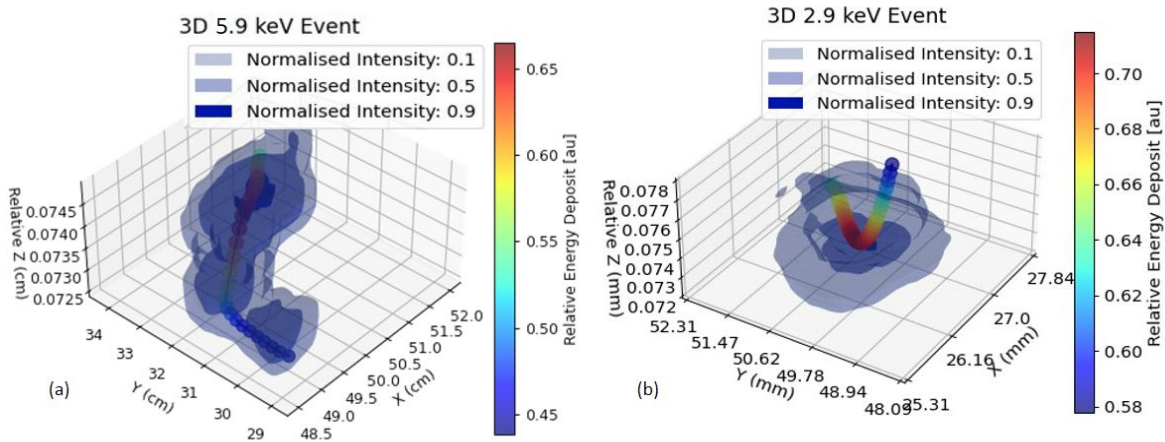

**Figure 3.** Examples of reconstructions from real data taken with the MIGDAL detector. Both reconstruction methods are represented in these plots. (a) is a 5.9 keV $^{55}$Fe event in $CF_4$. (b) is a 2.9 keV event consistent with the escape peak event in Ar.

## 3. Conclusion

The goal of the MIGDAL experiment is to detect and characterize the Migdal effect, which requires the 3D reconstruction of low dE/dx, low energy electron tracks. Using data obtained from the camera and strip readouts of the MIGDAL OTPC, we have demonstrated this for electron energies down to 2.6 keV using two different algorithms. While these results show promise for the MIGDAL experiment, they also demonstrate the potential of this technology for any application requiring full 3D reconstruction of low energy ionization tracks.

## Acknowledgments

This work has been funded by E. Tilly's U. S. Department of Energy Graduate Instrumentation Research Award (GIRA); by the U.S. Department of Energy, Office of Science, Office of High Energy Physics, under Award Number DE-SC0022357; by the U.S. National Science Foundation under Grant No. 2209307; UKRI's Science & Technology Facilities Council through the Xenon Futures R&D programme, Consolidated Grants, C. McCabe's Ernest Rutherford Fellowship and T. Marley's PhD scholarship; by the Portuguese Foundation for Science and Technology (FCT) under award number PTDC/FIS-PAR/2831/2020; and by the European Union's Horizon 2020 research and innovation programme under the Marie Sklodowska-Curie grant agreement No. 841261 (DarkSphere) and No. 101026519 (GaGARin).

---

[2] This success is determined qualitatively. Quantitative analysis on these reconstruction techniques is ongoing.